# Synthesis, characterization and modeling of high quality ferromagnetic Cr-doped AlN thin films

Stephen Y. Wu [a], H.X. Liu [a], Lin Gu [b], R.K. Singh [a], L. Budd [a], M. van Schilfgaarde [a], M.R. McCartney [b], David J. Smith [b,c], and N. Newman [1,a]

[a] Dept. of Chemical & Materials Engineering, Arizona State University, Tempe, AZ 85287-6006

[b] Center for Solid State Science, Arizona State University, Tempe, AZ 85287-1704

[c] Dept. of Physics and Astronomy, Arizona State University, Tempe, AZ 85287-1504.

[1] corresponding author

We report a theoretical and experimental investigation of Cr-doped AlN. Density functional calculations predict that the isolated Cr $t_2$ defect level in AlN is $\frac{1}{3}$ full, falls approximately at midgap, and broadens into an impurity band for concentrations over 5%. Substitutional $Al_{1-x}Cr_xN$ random alloys with $0.05 \leq x \leq 0.15$ are predicted to have Curie temperatures over 600 K. Experimentally, we have characterized and optimized the molecular beam epitaxy thin film growth process, and observed room temperature ferromagnetism with a coercive field, $H_c$, of 120 Oersted. The measured magnetic susceptibility indicates that over 33% of the Cr is magnetically active at room temperature and 40% at low temperature.

The combination of several discoveries in magnetic semiconductors has created a new field of research that has been labeled spintronics.[1] First, the observation of ferromagnetism in Mn-doped InAs and GaAs up to 100K has enhanced interest in finding and exploring semiconductors that exhibit ferromagnetism at even higher temperatures.[2] Second, the observation that spin-polarized populations exhibit long lifetimes and diffusion lengths in conventional semiconductors has created speculation that spintronics will lead to a number of new practical applications.[3] Third, a number of additional new semiconductor devices have been proposed that exploit ferromagnetism in semiconductors beyond the spin-polarized transistor first described by Datta and Das.[4] For example, manipulation of spin in semiconductor structures could be used to develop quantum-coherent electronics and possibly quantum computing.[3]

As we briefly describe here, and will describe in detail elsewhere[5], LSDA calculations predict that III-nitrides doped with small amounts of Mn or Cr will be ferromagnetic in the range of 500 to 800K. Bedair has reported room-temperature ferromagnetism in Mn-doped GaN grown by MOCVD.[6] Besides their high $T_c$, there are other reasons why the Cr-doped III-N system may be the material of choice for spintronic applications. First, III-N materials are expected to transport spin-polarized electrons over even longer times and distances than GaAs due to their reduced spin-orbit coupling. Second, other ferromagnetic III-V semiconductors produced to date (e.g. Mn-doped $Ga_{1-x}Mn_xAs$) contain high levels of compensation[7] and have poor crystalline quality as a result of the limited growth temperature necessary when depositing the volatile Mn. III-N materials are not expected to contain large levels of compensating defects. Furthermore, Cr can be incorporated at higher growth temperatures than Mn. Third, according to the LDA, partial compensation of Cr acceptors with donors can actually *enhance* $T_c$ because this can increase the filling of the defect band from $1/3$ towards the (optimal) ½ filling. Finally, calculations predict that for x ≥ 5%, Cr-doped GaN and AlN have higher $T_c$ than in the Mn-doped case.

There has been a recent report that polycrystalline AlN doped with Cr produced by sputtering is ferromagnetic.[8] However, less than ~5% of the dopants were reported to be magnetically active and the growth was done in poor vacuum conditions (i.e. $10^{-6}$ Torr base pressure), possibly resulting in the existence of trace amounts of $CrO_2$.

In this study, Cr-doped AlN films were synthesized on 6-H SiC(0001) substrates (Cree Research) in a custom-built reactive MBE system with a base pressure better than $5 \times 10^{-10}$ Torr. A description of the experimental methods can be found in Reference 11. The thickness and depth profile of the chemical composition have been characterized by Rutherford Backscattering Spectroscopy (RBS) with 2 and 3.05-MeV $He^{++}$ beams at incident angles of 8°. Characterization of magnetic properties was carried out using a superconducting quantum interference device (SQUID) magnetometer (S.H.E. VTS900). The microstructure of the Al(Cr)N films was characterized in the cross-sectional geometry using transmission electron microscopy (TEM) with a JEOL 4000EX high-resolution electron microscope operated at 400 keV.

Our LSDA calculations find that Mn and Cr, in the dilute doping limit, form near midgap deep levels in both zincblende (ZB) and wurtzite (WZ) AlN. Except were specified, the calculations employed 216-atom supercells in the ZB structure, and using the atomic spheres approximation (ASA). The *d* level is exchange-split, and in the ZB host are further split by the crystal-field split into a three-fold degenerate $t_2$ and a doubly degenerate *e* level, with the *e* level ~1eV below the $t_2$. The exchange splitting is ~2 eV. Levels in a WZ host are similar, but shifted upwards by ~0.2 eV. In AlN, presumably because of the rather large deviation from the ideal c/a ratio, the $t_2$ further splits in WZ into three distinct levels, separated by ~0.1 eV. The majority $t_2$ is $1/3$-filled for Cr and $2/3$-filled for Mn, leading to a magnetic moment of $3\mu_B$ for Cr and of $4\mu_B$ for Mn. As the concentration increases, these levels broaden into bands. The partial filling of this band accounts for the magnetism of a (substitutionally) doped $(Cr,Mn)_x(Ga,Al)_{1-x}N$ alloy. It is clear that the RKKY mechanism sometimes invoked to explain ferromagnetism in $Mn_xGa_{1-x}As$ does not apply to this case. To model the magnetic interactions, we employ the *ab-initio* LSDA, mapping the total energy onto a Heisenberg form.[12] There are no adjustable parameters, and moreover the theory encompasses the usual kinds of model forms of magnetic interactions typically invoked, such as the double-exchange and RKKY interactions. To estimate $T_c$ from the LDA Hamiltonian, a multicomponent mean-field theory was developed.[5] For the ideal Mn-doped GaN alloy, as in all ideal III-V DMS compounds studied to date, the magnetic exchange interactions as computed from the LDA linear-response eventually turn antiferromagnetic as the Mn concentration increases (see Fig. 1). Our mean-field analysis indicates that the $T_c$ is highest at about 10% Mn. The transition occurs at lower dopant concentrations in the N-bearing III-V's than when heavier anions (P, As) are used. But when doped with Cr, the exchange coupling and critical temperature (as estimated within a mean-field analysis of the LDA Hamiltonian) continue to increase with concentration, as shown in Fig. 1. Segregation of transition metal dopants can be a critical factor in determining and optimizing the FM properties.[10] The large exchange energy of the transition metal dopants provides a strong driving force for clustering.[10]

Initial growth studies consisted of varying deposition conditions in order to optimize the crystal quality and stoichiometry for a wide range of transition metal concentrations. Figure 2 shows the concentration of transition metal dopants incorporated as a function of substrate temperature. The transition metal dopant incorporation rate decreased as the substrate temperature increased due to the diminishing sticking coefficient at the higher temperatures. Also note that the incorporation rate of Mn dopants is significantly lower than that of Cr at a particular growth temperature.

Both the magnetic field dependence on magnetization, M, in the range of 10 to 300K, and the magnetic field dependence on temperature, in the range of 10 to 350K were measured with the external magnetic field parallel to the film plane. Figure 3 shows the magnetization as a function of applied external magnetic field at both 10K and 300K. These well-defined soft hysteresis loops confirm that the films are ferromagnetic at room temperature. We have observed spontaneous magnetization values, $M_s$, of 18.7 emu/cm$^3$ at 10K and 15.4 emu/cm$^3$ at 300K for $Al_{0.93}Cr_{.07}N_1$. These $M_s$ values indicate that the effective magnetic moment in Cr-doped AlN is 1.2 $\mu_B$/Cr atom at 10K and 1.0 $\mu_B$/Cr atom at 300K. Thus, over 33% of the Cr is magnetically active at room temperature and over 40% at low temperature. These values are approximately one order of magnitude larger than reported earlier.[8] The coercivity, $H_c$, is 120 Oe at room temperature. The inset in Figure 3 shows the temperature dependence on magnetization. Magnetization was kept constant at 0.1T and no sudden drop in magnetization was observed as the temperature was increased up to 350K, indicating a Curie temperature of greater than 350K. It is not clear if the less than 100% magnetic activity of the dopants can be attributed to the presence of a secondary phase or due to depletion of electrons from the $Cr_{Al}$ level by compensating acceptors.

From RBS analysis, the level of oxygen in the films is determined to be <5%, so Cr oxide is a possible secondary phase. While Cr and CrN are antiferromagnetic, Cr oxide is ferromagnetic, and could have been a source of the ferromagnetic signal observed. However, x-ray diffraction (XRD) showed no evidence of any Cr oxide phases, and small-probe microanalysis showed that the oxygen was uniformly distributed throughout the sample, whereas the Cr clearly showed strong signs of nanosegregation.

Cross-sectional electron micrographs indicated that the Al(Cr)N films generally had a distinct columnar morphology and the top surface of the layer had a facetted appearance, as shown in Fig. 4(a). Selected-area electron diffraction patterns (inset to Fig. 4(a)) showed excellent alignment of diffraction spots from the epilayer with those from the SiC substrate, indicative of overall high quality epitaxial growth. Small-probe microanalysis (FEI CM-200 field-emission-gun TEM) was used to determine the chemical composition throughout the epilayer. Figure 4(b) shows an annular-dark-field micrograph of the Al(Cr)N epilayer and Fig. 4(c) shows the x-ray spectra taken from along the line indicated. This line profile shows that the oxygen signal is low and relatively constant throughout the film whereas the Cr profile has comparatively large upward and downward swings, suggestive of Cr nanoclustering within the film. It is thus unlikely that any chrome-oxide phase, such as $CrO_2$ or $Cr_2O_3$, has been formed either during film deposition or subsequent annealing.

In summary, we have characterized the epitaxial growth of Cr-doped AlN thin films and observed room temperature ferromagnetism. The measured magnetic susceptibility indicates that over 33% of the Cr is magnetically active at room temperature and 40% at low temperature. According to the LSDA the ferromagnetism occurs in a midgap defect band in $Al_{1-x}Cr_xN$.

This work was supported by the Defense Advanced Research Projects Agency (DARPA) and administered by the Office of Naval Research (Contract No. N00014-02-1-0598, N00014-02-1-1025 and N00014-02-1-0967). The authors thank Professor Ralph Chamberlin for his assistance in bringing the SQUID magnetometer to fully operational capabilities.

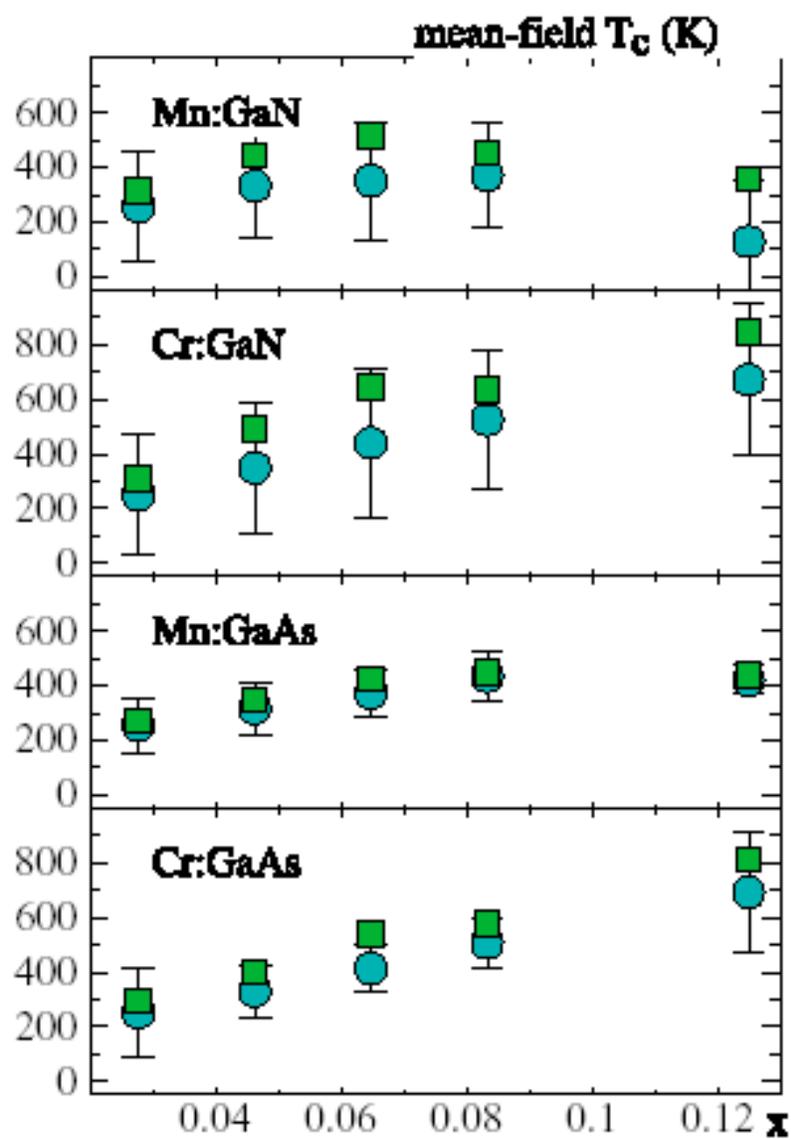

**FIG. 1. – Stephen Y. Wu, Applied Physics Letters**

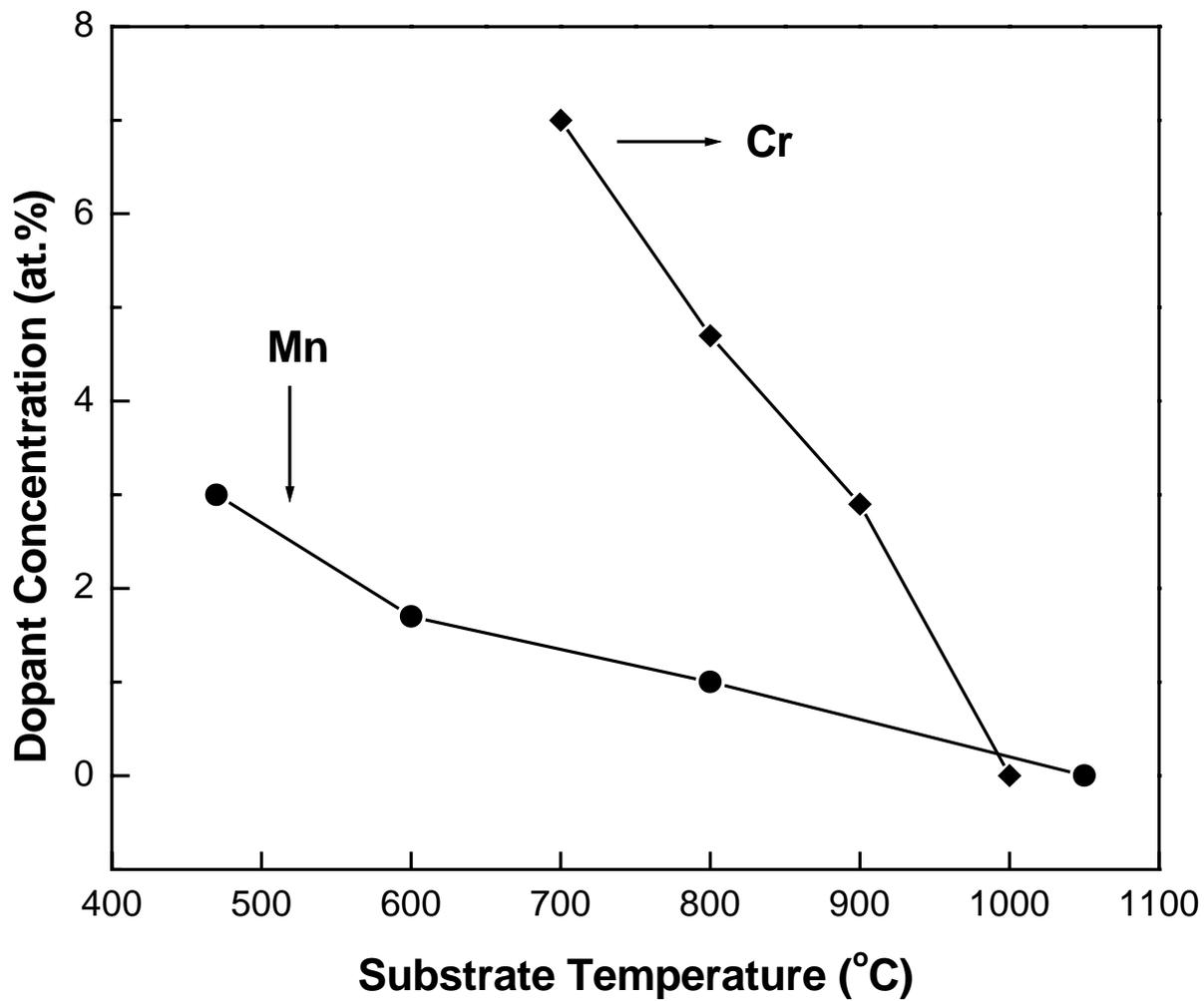

**FIG. 2.** – Stephen Y. Wu, Applied Physics Letters

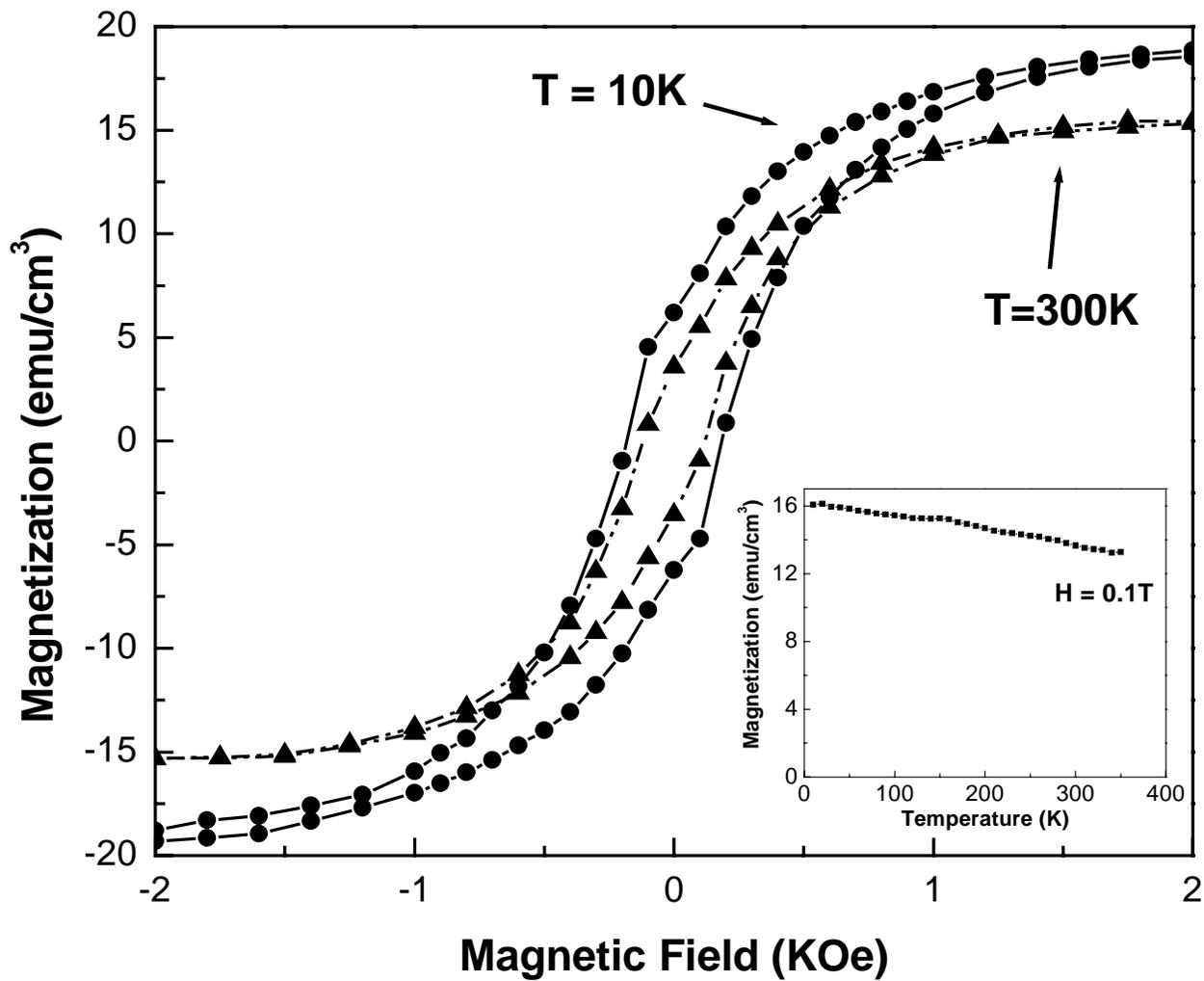

**FIG. 3 – Stephen Y. Wu, Applied Physics Letters**

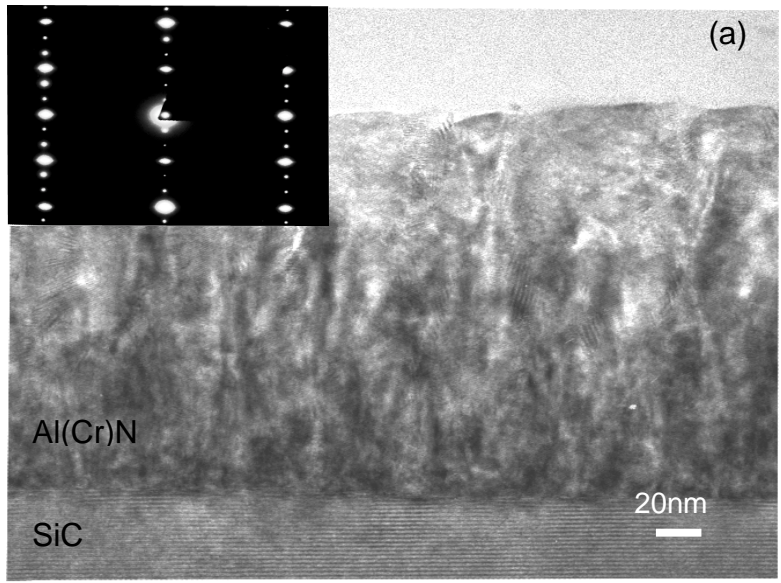

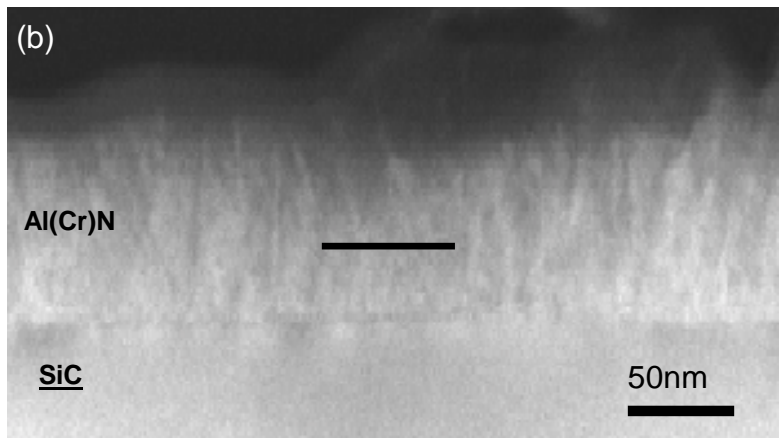

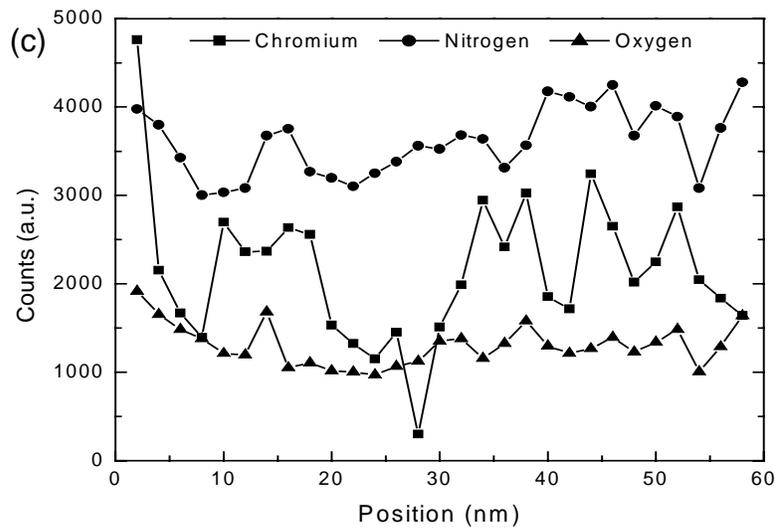

**FIG. 4. – Stephen Y. Wu, Applied Physics Letters**

Figure Captions

FIG 1. Mean-field estimates of $T_c$ computed from the LDA-derived heisenberg Hamiltonian, for a series of ideal random alloys in 216-atom zincblende supercells (for the 15.5% concentration a 128 atom special quasirandom structure was used). Because of the randomness of the magnetic atoms, there is a rather large fluctuation in the effective magnetic exchange field $J_0$ from one site to another. Circles show the average of $\frac{2}{3} J_0$ at each site, and error bars show the RMS fluctuations in $\frac{2}{3} J_0$. The quantity $\frac{2}{3} J_0$ is shown because for a homogeneous system (all magnetic atoms equivalent), $T_c = \frac{2}{3} J_0$ in mean-field theory. The squares show $T_c$ as computed by a multicomponent mean-field calculation. We find that the magnetic interactions for every configuration of Mn:AlN and Cr:AlN we studied (approximately half of those studied for GaN) are very similar to those of Mn:GaN and Cr:GaN, deviating from the GaN case by typically 10%.

FIG. 2. Transition metal dopant concentration incorporated into AlN as a function of substrate temperature.

FIG. 3. Magnetic field dependence of magnetization of AlN doped with 7% Cr at 10K and 300K. Inset shows temperature dependence on magnetization kept constant at 0.1T.

FIG.4. (a) Cross sectional TEM micrograph of Cr-doped AlN film showing distinct columnar morphology. Inset shows a selected-area electron diffraction pattern representative of high quality epitaxial growth. (b) Small-probe microanalysis. Annular-dark-field micrograph of Al(Cr)N epilayer. (c) X-ray spectra taken along line shown in (a). Oxygen is relatively uniform throughout and signal fluctuation of Cr is indicative of Cr nanoclustering.